\documentclass[aps,pre,twocolumn,showpacs,floatfix,superscriptaddress]{revtex4-1}
\usepackage{graphicx}
\usepackage{amssymb}
\usepackage{color}
\usepackage{amsmath}
\usepackage{ulem}
\usepackage{mathrsfs}

\begin{document}

\title{Dynamic scaling in the two-dimensional Ising spin glass with normal-distributed couplings} 
  
\author{Na Xu}
\affiliation{Department of Physics, Boston University, 590 Commonwealth Avenue, Boston, Massachusetts 02215, USA}

\author{Kai-Hsin Wu}
\affiliation{Department of Physics and Center of Theoretical Sciences, National Taiwan University, Taipei 10607, Taiwan}

\author{Shanon J. Rubin} 
\affiliation{Department of Physics, Boston University, 590 Commonwealth Avenue, Boston, Massachusetts 02215, USA}

\author{Ying-Jer Kao}
\affiliation{Department of Physics and Center of Theoretical Sciences, National Taiwan University, Taipei 10607, Taiwan}
\affiliation{National Center of Theoretical Sciences, National Tsinghua University, Hsinchu, Taiwan}

\author{Anders W. Sandvik}
\affiliation{Department of Physics, Boston University, 590 Commonwealth Avenue, Boston, Massachusetts 02215, USA}

\begin{abstract}
We carry out simulated annealing and employ a generalized Kibble-Zurek scaling hypothesis to study the two-dimensional Ising spin glass with 
normal-distributed couplings. The system has an equilibrium glass transition at temperature $T=0$. From a scaling analysis when 
$T\rightarrow 0$ at different annealing velocities $v$, we find power-law scaling in the system size for the velocity required in order 
to relax toward the ground state; $v \sim L^{-(z+1/\nu)}$, the Kibble-Zurek ansatz where $z$ is the dynamic critical exponent and $\nu$ 
the previously known correlation-length exponent, $\nu\approx 3.6$. We find $z \approx 13.6$ for both the 
Edwards-Anderson spin-glass order parameter and the excess energy. This is different from a previous study of the system with bimodal 
couplings [S. J. Rubin, N. Xu, and A. W. Sandvik, Phys. Rev. E {\bf 95}, 052133 (2017)] where the dynamics is faster ($z$ is smaller)
and the above two quantities relax with different dynamic exponents (with that of the energy being larger). We argue that the different 
behaviors arise as a consequence of the different low-energy landscapes---for normal-distributed couplings the ground state is unique 
(up to a spin reflection) while the system with bimodal couplings is massively degenerate. Our results reinforce the conclusion of 
anomalous entropy-driven relaxation behavior in the bimodal Ising glass. In the case of a continuous coupling distribution, our results 
presented here also indicate that, although Kibble-Zurek scaling holds, the perturbative behavior normally applying in the slow limit 
breaks down, likely due to quasi-degenerate states, and the scaling function 
takes a different form.
\end{abstract}

\date{\today}

\maketitle

\section{Introduction}

Spin glasses are benchmark models for studying complex physical systems and optimization problems. Due to the disorder and frustration (random mixed-sign 
couplings), the energy landscapes of these systems are very rough, with many local minimums, and it is very challenging to find the true global minimum (ground 
state) through Monte Carlo (MC) simulations \cite{edwards75,binder86,fisher91,parisi83}. Among the common spin glass systems, the two-dimensional (2D) Ising spin glass (2DISG) 
is special in that the paramagnetic--glass phase transition occurs exactly at temperature $T=0$. The system has long-range spin-glass order (defined
with the Edwards-Anderson, EA, order parameter) at $T=0$, and the correlation length diverges as a power law, $\xi\sim T^{-\nu}$ when $T\rightarrow 0$.  
Many works have been devoted to the nature of the critical behavior and to obtaining the critical exponents of 2DISG system with both normal-distributed
(Gaussian) and bimodal couplings \cite{kawashima92,campbell04,houdayer04,rieger96}. However, due to the considerable challenges with MC simulations, 
especially for large systems at low temperature, there are still significant issues under debate. For example, whether or not the 2DISG 
with bimodal $J = \pm 1$ and Gaussian couplings belong to the same universality class in their equilibrium criticality is still in question \cite{jorg06,toldin11,thomas11,fernandez16,campbell16,campbell17}. Undisputed is the fact that the ground-state properties of the two models are different. 
The system with Gaussian couplings has a unique (non-degenerate) ground state, up to a trivial spin reflection, while the model with bimodal couplings 
has infinite degeneracy in the thermodynamic limit. 

Given the difficulties in studying the critical behavior through equilibrium simulations, the recently developed non-equilibrium approach based on generalized
Kibble-Zurek (KZ) scaling \cite{kibble76,zurek85,zhong05,chandran12,liu14,liu15a,polkovnikov05,zurek05} provides a powerful alternative method 
for studies of spin-glass models. KZ scaling of simulated annealing (SA) results has been successfully applied to 3D and 2D spin glass systems 
in order to extract the dynamic exponent $z$ and other critical exponents \cite{liu15,rubin17}. The key aspect of the KZ mechanism used in this context is 
the prediction that slow (close to equilibrium) and fast (far from equilibrium) SA processes are separated by an annealing velocity $v_{\rm KZ}$ that scales
with the system size (length) $L$ as
\begin{equation}
v_{\rm KZ} \propto L^{-z-1/\nu},
\label{vkzdef}
\end{equation}
where $\nu$ is the standard equilibrium correlation-length exponent.
Here we apply this approach to the 2DISG with Gaussian couplings, following the recent work on bimodal couplings \cite{rubin17}.

In Ref.~\onlinecite{rubin17}, a surprising behavior with dual 
time scales governing the relaxation when $T\to 0$ was discovered. Contrary to the general expectation that the order parameter is the slowest-relaxing
physical observable, and that most other quantities are asymptotically governed by that same time scale, a larger dynamic exponent, $z_{E}\approx 10.3$,
was found for the excess energy than $z_{q}\approx 8.3$ for the EA order parameter. The physical mechanism proposed to underly the two time scales 
relies on the backbone (largest common cluster) and droplet (zero-energy flippable cluster) structure of the massively degenerate ground states of the $J=\pm 1$ 
model \cite{roma10,thomas11}, which leads to a concentration in the configuration space of low-energy states that entropically attracts the SA process.
The proximity of true ground states and low-energy excitations to each other within this region was proposed to lead to an insensitivity of the replica-overlap 
definition of the order parameter to low-energy excitations, so that the final relaxation of the energy leads to only sub-leading corrections to the already 
equilibrated mean order parameter. This $T \to 0$ relaxation process is of particular relevance in related optimization problems, where currently there
is much interest in comparing SA and quantum annealing protocols and the connectivity (especially the dimensionality) of the the spins (qubits 
in quantum annealing) may play a very important role \cite{katzgraber15}.

It should be noted that the relaxation dynamics in an SA process for $T \to 0$ can be very different from the dynamics associated with ergodic sampling at 
fixed $T >0$. The latter should be associated with a divergent dynamic exponent when $T\to 0$ in 2D Ising spin glasses \cite{katzgraber05}, which
also is consistent with the non-ergodicity of local spin moves at $T=0$. In SA, the temperature is constantly changing and naive arguments based on activated
dynamic scaling to overcome energy barriers do not necessarily apply in all cases, since the details of the energy landscape matter. In Ref.~\onlinecite{rubin17}
it was argued that the droplet structure of the 2D Ising spin glass corresponds to a funnel-like feature of the energy landscape where high energy
barriers can be overcome at high temperatures and the barriers remaining as the ground state is approached when $T\to 0$ become typically smaller, such that 
a power-law scaling of the annealing time required to reach the ground state is obtained. This situation is also of great interest in the context of 
optimization and computational complexity, as a case where the typical exponential scaling to find an optimal solution can be avoided \cite{katzgraber15}.

In the case of Gaussian-distributed couplings, which we study in this paper, the backbone structure can be defined only as an approximation with low-energy states 
instead of true ground states \cite{roma13,roma16}. Strictly speaking, there is no definable backbone and zero-energy clusters in that model due to the lack of 
ground-state degeneracy. Because of this qualitative difference of the ground-state landscape, one can expect different dynamical properties for the Gaussian 
model (or any other continuous coupling distribution). The aim of the work presented here is to apply exactly the same scaling approach as was done with the 
bimodal 2DISG in Ref.~\onlinecite{rubin17} and test whether a clearly different asymptotic relaxation mechanism can be detected. We will show that, indeed, in this 
case the excess energy and the EA order parameter relax with the same dynamic exponent, and the value of the exponent, $z =13.6(2)$ (where here and later the
number in parentheses indicates the statistical error of the preceding digit), is significantly larger than both exponents found in the bimodal case.

The organization of the rest of the paper is as follows: In Sec.~\ref{sec:methods} we discuss the known equilibrium properties and expected finite-size 
behaviors near the $T=0$ critical point of the 2DISG model with Gaussian couplings. These properties are important when extending finite-size scaling to
non-equilibrium setups where the annealing velocity enters as another variable. We describe the SA simulation procedures, where we have applied GPU 
(graphics processing-unit) computing for very efficient MC sampling with the Metropolis algorithm, and summarize the KZ scaling procedures 
we have applied to quantify the relaxation behavior as a function of system size and annealing velocity. In Sec.~\ref{sec:re} we present results of the
scaling analysis for the excess energy and the EA order parameter. Lastly, in Sec.~\ref{sec:dis} we further discuss our findings and contrast them with 
the conclusions previously drawn for the bimodal case. 

\section{Model and methods}
\label{sec:methods}

The Hamiltonian of the 2DISG is
\begin{equation}
H = \sum_{\langle  ij\rangle }J_{ij} \sigma_{i} \sigma_{j},~~~ \sigma_i=\pm 1,
\label{ham}
\end{equation}
where, in the case considered here, $\langle  ij\rangle $ stands for nearest-neighbour spins on a 2D square lattice with $L^2$ sites and 
periodic boundary conditions. The couplings $J_{ij}$ are drawn from some distribution, here Gaussian with mean $0$ and standard deviation $1$.

\subsection{Equilibrium finite-size scaling}
\label{subsec:fss}

The primary quantity capturing the spin-glass phase transition is the EA order parameter,
\begin{equation}
q = \frac{1}{N}\sum_{i=1}^N \sigma^{(1)}_i\sigma^{(2)}_i,
\label{ea}
\end{equation}
where (1) and (2) stand for two independently generated configurations (two different MC simulations), also referred to as `replicas', of systems with 
the same coupling realization $\{J_{ij}\}$. In this paper, we focus on the mean squared EA order parameter, $\langle  q^2\rangle$, as well as 
the internal energy density $E=\langle  H\rangle /N$ in the limit $T\to 0$ reached in SA simulations with Metropolis dynamics. For simplicity 
of notation, we use $\langle ...\rangle$ to denote the combined MC expectation value and the average over disorder samples.

It is known that the 2DISG with Gaussian couplings has a phase transition exactly at $T=0$, and its critical behavior has been studied extensively
\cite{kawashima92,campbell04,houdayer04}. Unlike the 2DISG with $J=\pm1$ couplings, where there are many degenerate ground states, there is only a 
unique ground state (and the state with all spins reversed). Thus, as $T\rightarrow0$ all the independent replicas will eventually fall into the same ground 
state configuration in the limit of a very slow SA process, and the EA order parameter $\langle q^2\rangle$ must approach $1$ without any finite-size 
corrections in the $T=0$ value. However, according to the study in \cite{campbell04}, the equilibrium ground-state energy density has a finite-size correction
of the form 
\begin{equation}
E(L)-E_{\infty} = aL^{-(d+\frac{1}{\nu})}.
\label{de}
\end{equation}
Here the energy per spin for infinite $d=2$ dimensional system is $E_{\infty}=-1.314788(4)$ \cite{thomas07} and the most precise value available for
the critical exponent $\nu$ of the correlation length $\xi$ (where in the case $T_c=0$ we have $\xi \sim T^{-\nu}$) is $\nu=3.56(2)$ \cite{campbell04}.
The prefactor $a$ of the scaling in $L$ was claimed to be exactly $a=1$. In the following analysis of SA data, we will make use of the 
form (\ref{de}) with the previously determined values of $E_{\infty}$ and $\nu$ (while the value of $a$ is less important).

\subsection{Simulated annealing}
\label{subsec:sa}

Most of the simulations were run on Nvidia CUDA enabled GPUs, with single-spin  Metropolis updates and multi-spin coding where the Ising spins 
$\sigma_i = \pm 1$ of the model [Eq.(\ref{ham})] are coded as bits of 32-bit integers. Thus, with the same set of random couplings, one simulation propagates 
$32$ replicas from different initial conditions and the order parameter $q^2$ is computed at the end of the run (at $T=0$) from the overlap, of the form
Eq.~(\ref{ea}),  among these replicas. One sweep of MC updates involves $N=L^2$ Metropolis spin-flip attempts, carried out successively in two groups corresponding 
to the standard checker-board decomposition of the lattice, so that all spins in a given sublattice can be updated in parallel independently of each other.
For each of the 32 replicas, different random numbers are generated in order for the updating processes to be fully independent. Our program achieves
around $2\times 10^{9}$ attempted spin flips per second on a single GPU. Further discussion on how to implement these updates on GPUs can be found in
Refs.~\onlinecite{preis09,block10,lulli15,hsieh13,hartmann15,fang14,jesi14,parisi10,weigel12a,weigel12b}. 

In an SA procedure, after initial equilibration of the system at a high temperature $T_{\rm ini}$, the temperature $T(t)$ is lowered as a function of the
simulation time $t$ according to some protocol. In the context of the KZ mechanism one normally considers the approach to a phase transition using a linear 
protocol, or, if the transition point is known one can approach it with a nonlinear power-law protocol (or in principle some other protocol).
Note that we are here not interested in finding an optimal SA protocol (i.e., the one that would bring us to the ground state in the shortest time), 
but aim to test the power-law KZ scaling hypothesis for the 2D Ising glass with transition temperature exactly at $T=0$, and use it to extract dynamic 
information.

The general power-law protocol takes us to $T=0$ as a function of the total simulation time $t_{\rm max}$ according to
\begin{equation}
T(t) = T_{\rm ini}(1-t/t_{\rm max})^r.
\label{tprotocol}
\end{equation}
Here $r=1$ corresponds to the standard SA protocol where the temperature is decreased linearly.  In order to disentangle the exponents $z$ and $\nu$ involved 
in KZ scaling, e.g., in Eq.~(\ref{vkzdef}),  it is also useful to study other values of $r$, as exemplified in the previous study of the 2DISG with bimodal
couplings \cite{rubin17}. There a consistency check was provided by the fact that the entropy exponent $\Theta_S$, which plays the role of $1/\nu$ in that
case \cite{thomas11}, was determined independently and agreed with previous calculations. In the work reported here, we only consider $r=1$ and use the known
value of $\nu$ to extract $z$, because the
calculations with Gaussian couplings are very expensive (even with the use of GPUs), mainly due to the fact that longer times are needed to reach close to the 
unique ground state. Another reason for only considering $r=1$ is that the value of $1/\nu$ is small, around $0.28$, and it is then hard to determine it 
independently from simulations of two or more $r$ values within the error levels we can reach for the KZ exponent, $z+1/\nu$ in Eq.~(\ref{vkzdef}) for
$r=1$ [and $z+1/r\nu$ for other $r>0$ \cite{polkovnikov05}, which we do not consider here]. 

The annealing velocity is defined as $v=T_{\rm ini}/t_{\rm max}$. At the last step of the 
annealing process, when $T$ has 
reached $0$, we take measurements of the EA order parameter $q^2$, the energy (per site) $E$ of the system, as well as the minimum energy per spin, $E_{\rm min}$, 
among any of the $32$ replicas. The SA process is repeated many times for different realizations of the random couplings.

To test the SA program, for $L \le 6$ we used sufficiently long simulations for several disorder realizations to relax the systems all the way to the 
ground state. We checked these results against exact ground states, which can be obtained using an exhaustive search in the state space or using a matching 
algorithm such as those described in Refs.~\onlinecite{thomas07,hartmann11}. Based on the tests we know that the ground states are indeed reached in the simulations 
for sufficiently slow $v$, as expected. For the mean values taken over a large number of samples that we report below, because of the long simulation times
we were not able to use low enough $v$ for even the small-$L$ systems to reach their ground state in all cases, and for larger $L$ none of the systems reached 
the ground state. As we will see below, the mean values still do reach sufficiently close to their ground-state values to test the asymptotic KZ relaxation behavior.

\subsection{Dynamic scaling}
\label{subsec:ds}

In a generalized KZ scaling ansatz, for a system reaching the critical point through the annealing protocol expressed in Eq.~(\ref{tprotocol}), a physical 
quantity $A$ evaluated at the critical point can be written in the following finite-size scaling form \cite{zhong05,chandran12,liu14}:
\begin{equation}
A(v,L)=A_{\rm eq}(L) f(v/v_{\rm KZ}),
\label{avl}
\end{equation}
where the ``critical'' KZ velocity for the linear SA protocol, $r=1$ in Eq.~(\ref{tprotocol}), is given by Eq.~(\ref{vkzdef})
up to an undetermined and essentially arbitrary factor. This velocity demarks the borderline between fast and slow annealing processes. The function
$A_{\rm eq}(L)$ in Eq.~(\ref{avl}) stands for the equilibrium finite-size dependent quantity $A$ at the critical point, which normally is a power of $L$ 
to leading order but also can include scaling corrections. The dynamic exponent relates the scaling of the relaxation time 
to the correlation length through $\tau\sim\xi^z$, which at the critical point for finite-size systems turns into $\tau\sim L^z$ by the standard
substitution $\xi \to L$ in critical finite-size scaling. Recall the discussions in Sec.~\ref{subsec:fss}, that when the system is in equilibrium at 
$T=0$ the order parameter $\langle q^2\rangle =1$ without any finite-size effect, while the excess energy density has a finite-size correction of 
the form Eq.~(\ref{de}). These behaviors will be reflected in the corresponding $A_{\rm eq}(L)$ in Eq.~(\ref{avl}).

According to the general non-equilibrium scaling form that describes the dynamics in its full regime of velocities and sufficiently large system 
sizes, the order parameter $\langle q^2\rangle $ can be written in the following way \cite{liu14}:
\begin{equation}
 \langle  q^2(v,L)\rangle  \propto \begin{cases}
     f_0(vL^{z+1/\nu}),~~  v \alt v_{\rm KZ},\\
     \\
     (vL^{z+{1}/{\nu}})^{-x},~~  v_{\rm KZ} \alt v \alt 1,\\
     \\
     L^{-2}f_1(1/v),~~  v \agt 1. \label{eq:q_scaling}
     \end{cases}
\end{equation}
Here the first line describes the slow velocity regime, where the function $f_0$ normally would be a regular (Taylor-expandable) function of the 
KZ-scaled velocity $vL^{z+1/\nu}$, although below we will argue that, in the case considered here, a corresponding function of a power of the KZ variable 
has to be used for this to be true.  As discussed above, there should not be any dependence on the system size asymptotically for $v \to 0$ since 
$\langle q^2\rangle \to 1$ because of the unique ground state. 
In principle there could be $L$-dependent corrections for $v>0$ but the 
form of these are not presently known. The third line describes the fast velocity regime, in which the system size is larger than the correlation 
length $\xi_v$ at the end of the annealing process and, thus, there is no dependence on $L$, other than the trivial factor $L^{-2}$ that follows
from Eq.~(\ref{ea}) when the spin-glass correlation length is finite. 
The function $f_1$ should be Taylor-expandable in $1/v$. The second line in Eq.~(\ref{eq:q_scaling}) describes the 
intermediate power-law regime that connects the two other regimes. It follows from the scaling hypothesis, Eqs.~(\ref{avl}) and (\ref{vkzdef}), where 
the behavior when $L \to \infty$ at fixed $v$ must reduce to (connect smoothly to) the form on the third line, again because $\xi_v \ll L$ in this limit.
The only way to make this possible (i.e., to ensure the same $L$-dependence in the two forms) is with the power-law form 
$f(v/v_{\rm KZ}) \to $ $(vL^{z+{1}/{\nu}})^{-x}$, where the exponent $x$ must be given by
\begin{equation}
x=\frac{2}{z+1/\nu},
\label{x_q}
\end{equation}
so that the power-law form can also be written as $\langle q^2\rangle \propto L^{-2}v^{-x}$. 
Then the connection between lines 2 and 3 in Eq.~(\ref{eq:q_scaling}) corresponds to the function $f_1(1/v)$ crossing 
over into the form $v^{-x}$ and the connection between lines 1 and 2 corresponds to $f_0(vL^{z+1/\nu})$ taking the form
$(vL^{z+{1}/{\nu}})^{-x}$ for large $vL^{z+{1}/{\nu}}$. In other words, the KZ scaling form in Eq.~(\ref{avl}), with the KZ 
velocity given by Eq.~(\ref{vkzdef}), covers the first and second lines of Eq.~(\ref{eq:q_scaling}), while the third line represents 
the break-down of this form for higher velocities.

We also consider the  excess energy density, which we here define relative to the known infinite-size equilibrium $T=0$ value $E_\infty$,
\begin{equation}
\Delta E(v,L) =E(v,L)-E_{\infty},
\label{dedef}
\end{equation}
i.e., it contains contributions from both finite size and non-zero velocity. In analogy with the above discussion of the EA order parameter, 
and considering the equilibrium finite-size scaling given in Eq.~(\ref{de}), the behaviors in the three different velocity regimes should be
given by
\begin{equation}
 \langle  \Delta E(v,L)\rangle  \propto \begin{cases}
     L^{-(2+1/\nu)}g_0(vL^{z+1/\nu}),~~  v \alt v_{\rm KZ},\\
     \\
     L^{-(2+1/\nu)}(vL^{z+{1}/{\nu}})^{-x'},~~  v_{\rm KZ} \alt v \alt 1,\\
     \\
     g_1(1/v),~~  v \agt 1, \label{eq:e_scaling} 
     \end{cases}
\end{equation}
where, unlike Eq.~(\ref{eq:e_scaling}), there is no $L$ dependence on the third line because the excess energy is defined per spin
and takes a constant value when $v\to \infty$ (i.e., in the initial state). In this case, for the power-law regime to be valid, i.e., 
for there to be no size dependance on the second line ($\Delta E \sim v^{-x'}$), the exponent $x'$ is given by
\begin{equation}
x'=\frac{2+1/\nu}{z+1/\nu}.
\label{x_E}
\end{equation}
In the next section, we will present our results of the application of the above scaling forms.

\section{Results}
\label{sec:re}

All simulations reported here started from $T_{\rm ini}=8$, where the system can be easily equilibrated. Starting from a random configuration for each 
disorder sample, we used 10 MC sweeps at this initial temperature. From there, we used the linear SA process, i.e., $r=1$ in Eq.~(\ref{tprotocol}),
and measurements were taken at the last step of the annealing process where $T=0$. We used system sizes from $L=4$ to $L=64$. To span a wide range of 
velocities, we take the total time for the simulations as $t_{\rm max}=2^n$, where $n=2,3,...,30$ for small system sizes, while for large system sizes 
we only used $n$ up to $28$ to stay within reasonable computing times. To obtain good statistical averages, we simulated at least $5\times 10^3$ coupling 
realizations in most cases and $10^3$ realizations for the lowest velocities and largest system sizes. 

\subsection{Mean excess energy density}

\begin{figure}[t]
\center{\includegraphics[width=7cm, clip]{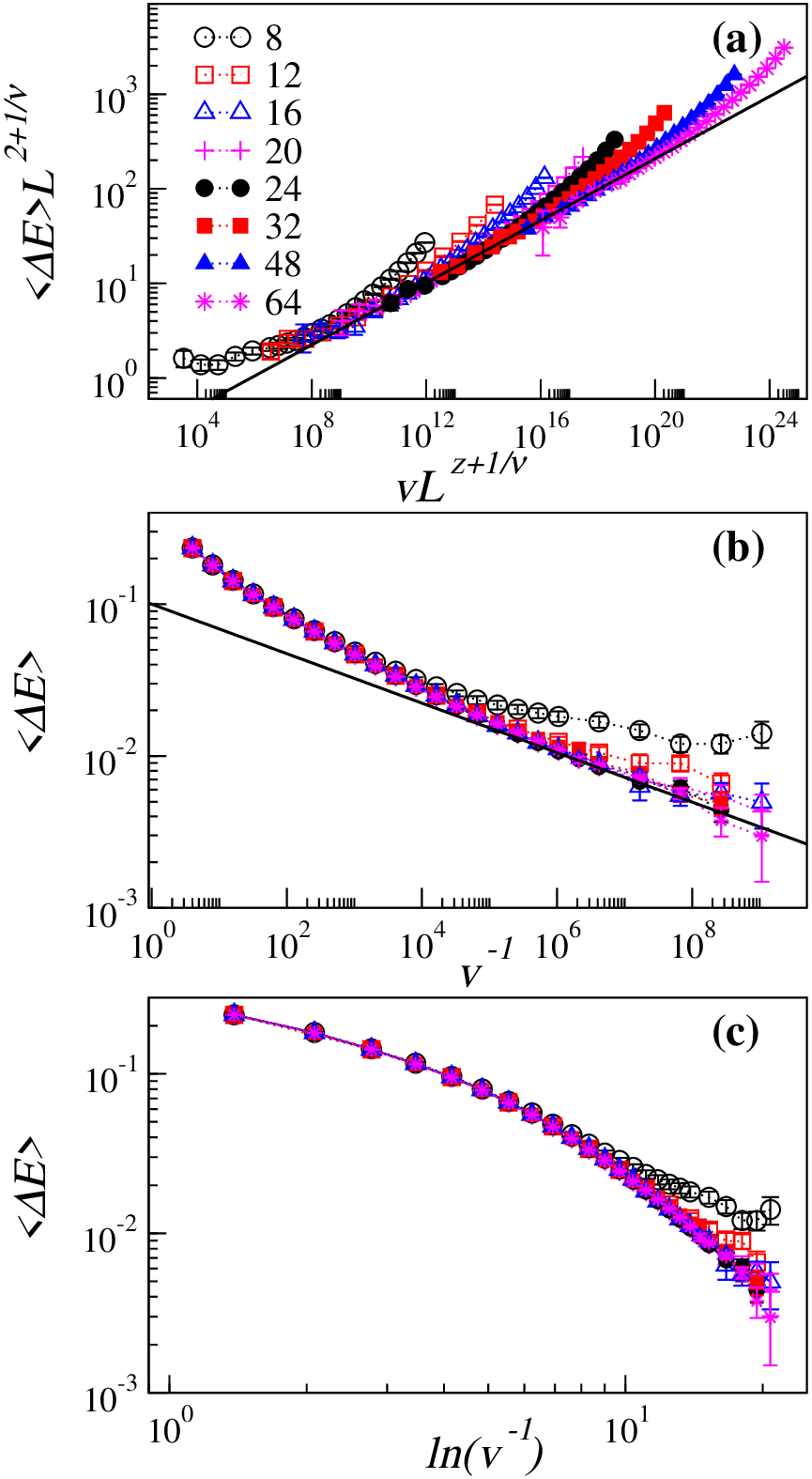}}
\vskip-1mm
\caption{(a) Velocity scaling of the mean excess energy density, $\Delta E=E(v,L)-E_{\infty}$. The data collapse for system sizes in the range $L=8$ to $64$ 
is optimal for $z=13.6(4)$. The straight line indicates the power-law regime with the expected exponent $x'$ given by Eq.~(\ref{x_E}). (b) The same data 
graphed according to the third line of Eq.~(\ref{eq:e_scaling}). The line shows the expected power-law behaviour with exponent $-x'$. (c)
The data  graphed versus $\ln(1/v)$.} 
\label{fig01}
\vskip-1mm
\end{figure}

Figure~\ref{fig01}(a) shows the velocity scaling of the average of the  excess energy density, Eq.~(\ref{dedef}), with $E_{\infty}=-1.31479$ from 
Refs.~\onlinecite{campbell04,thomas07}. The overall expected size dependence in equilibrium from Eq.~(\ref{de}) has been divided out, and the velocity 
has been rescaled according to the expected KZ form in Eqs.~(\ref{avl}) and (\ref{vkzdef}). Here we use data points from system sizes $L=8$ to $L=64$ in the 
data-collapse procedure, for each $L$ excluding velocities too high to give results on the common scaling function. We vary the scaling exponent 
$z+1/\nu$ to achieve optimal collapse relative to a fitted polynomial, repeating the procedure many times with Gaussian noise added to the data points 
in order to compute the statistical error. We obtain $z+1/\nu=13.9(4)$. Since $\nu \approx 3.56$ \cite{campbell04}, the dynamic exponent governing 
the excess energy is $z=13.6(4)$. 

In Fig.~\ref{fig01}(a) it is clear that data points for larger $v$ systematically peel off from the collapsed function and the region of data collapse 
in the rescaled variable is pushed further out to the right as $L$ increases. Figure \ref{fig01}(b) shows the same data graphed according to 
the third line of Eq.~(\ref{eq:e_scaling}). The data now collapse well for high velocities, and instead the data for slower velocities peel off systematically
from the common function as equilibrium is approached for each system size (i.e., the correlation length $\xi_v$ becomes of the order of the system size). In 
both Fig.~\ref{fig01}(a) and Fig.~\ref{fig01}(b), the straight lines indicate power-law behavior as described in the second line of Eq.~(\ref{eq:e_scaling}) 
with the expected slopes, $x'$ and $-x'$, respectively, given by Eq.~(\ref{x_E}). 

In order to test alternatives to the KZ scenario, we have also analyzed the data in other ways. One might naively expect that the SA relaxation of the system
should involve an exponentially long time scale when $T\to 0$, given energy barriers that have to be overcome when the system becomes trapped in local energy
minimums. The resulting activated scaling is reflected in a divergent {\it equilibrium} dynamic exponent $z_{\rm eq}(T)$ when $T\to 0$ \cite{katzgraber05}.
In Fig.~\ref{fig01}(c) we test for activated scaling of SA in the thermodynamic limit by graphing the same data as in Fig.~\ref{fig01}(b) versus $\ln(1/v)$ instead 
of $1/v$, still using logarithmic scales on both axes. On this plot a linear dependence would imply $\Delta E \sim \ln^{-a}(1/v)$ with some positive exponent
$a$, instead of the behavior $\Delta E \sim v^{x\prime}$ that we argued for above. We do not see any clear-cut linear behavior on the log-log plot, with
more curvature in the system-size converged data for the lowest velocities than in the KZ-scaled data in Fig.~\ref{fig01}(b). While one could perhaps
argue that the data approach a straight line also here, we point out that the KZ form $\Delta E \sim v^{x\prime}$ with a small exponent $x\prime \approx 0.17$
will inevitably look similar to the form $\Delta E \sim \ln^{-a}(1/v)$ in a limited window of the argument $\ln(1/v)$, because a small power looks very similar
to a logarithm. Thus, if in the window in question we have $\ln(1/v) \sim v^{-b}$ for some small value of the exponent $b$, then the KZ form will look
like $\Delta E \sim \ln^{-x\prime/b}(1/v)$, so that the exponent $a$ above is roughly $x\prime/b$. 

\begin{figure}[t]
\center{\includegraphics[width=7cm, clip]{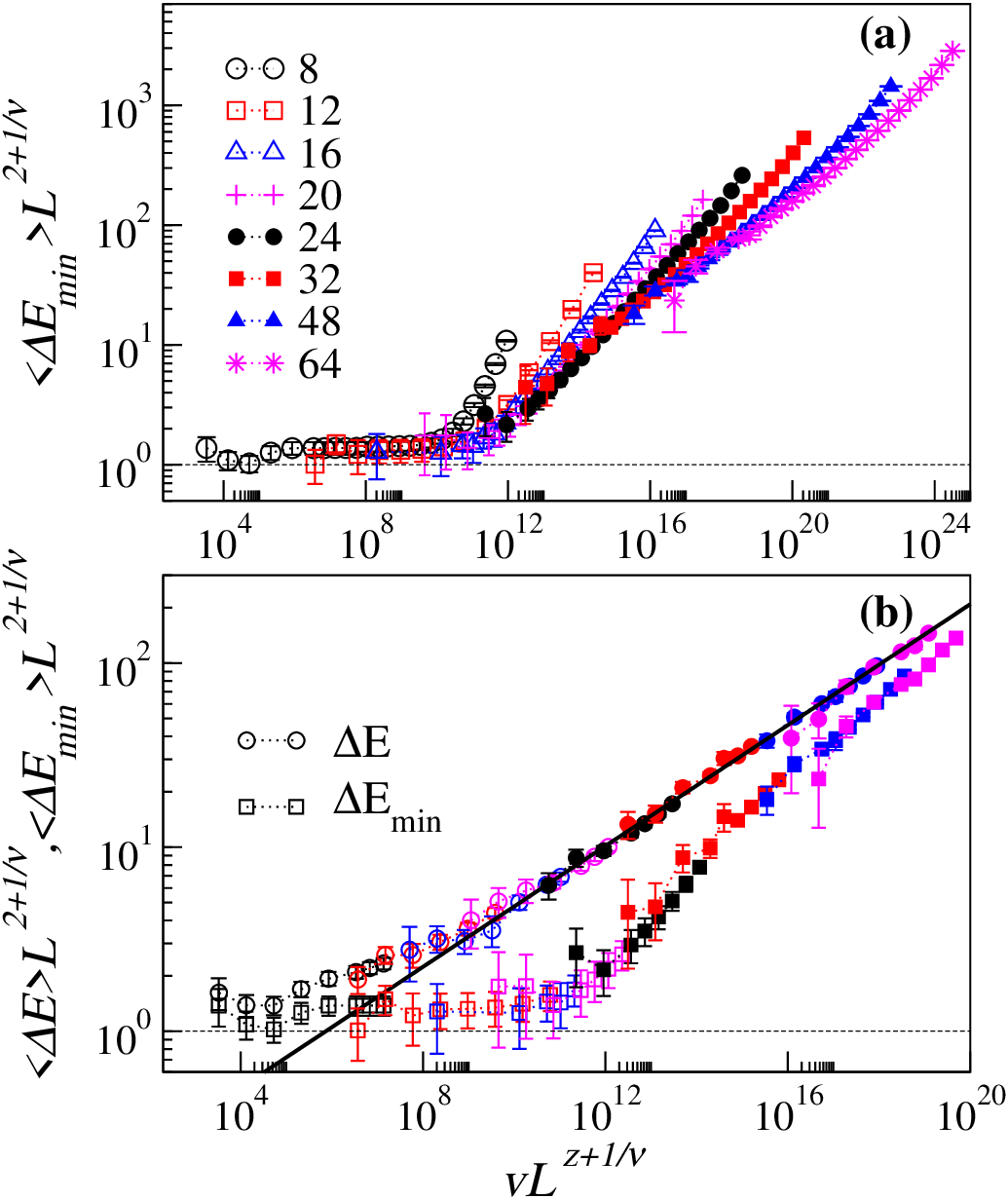}}
\vskip-1mm
\caption{(a) Velocity scaling of the minimum excess energy $\Delta E_{\rm min}$ per spin, where the exponent $z+1/\nu=13.9$ is the same as in Fig.~\ref{fig01}(a). 
(b) Scaling of both $\Delta E$ and $\Delta E_{\rm min}$, with the same exponent as in (a) and only including the well-collapsed data in order to make the
scaling functions better visible. The straight line is the same as in Fig.~\ref{fig01}(a). In both panels, the asymptotic value of the scaled quantities
for small $vL^{z+1/\nu}$ is consistent with the coefficient $a=1$ in Eq.~(\ref{de}), as indicated by the dashed lines.}
\label{fig02}
\vskip-1mm
\end{figure}

Note again that the KZ scaling demonstrated in panels (a) and (b) of Fig.~\ref{fig01} is not merely relying on the power-law scaling in the limit $L\to 0$ in
a rather small window of velocities that we have achieved, but is mainly manifested in the generalized finite-size scaling form that applies also when equilibrium 
is reached for the smaller system sizes in panel (a). Importantly, there is full consistency of the asymptotic slope in panel (b) with the exponent $x'$ defined in 
Eq.~(\ref{x_E}) with the value of $z$ that also describes the data collapse to the left of the power-law regime in Fig.~\ref{fig01}(a), i.e., the KZ scaling 
hypothesis also describes the deviations from the infinite-size collapsed curve for the smaller system sizes ($L=8$ and $12$) in Fig.~\ref{fig01}(b).
In combination with the previous results for the bimodal coupling distribution in Ref.~\onlinecite{rubin17}, where the dynamic exponent is smaller 
and the KZ behavior can be seen even more clearly, we take these results as strong evidence of KZ scaling also with the normal-distributed couplings.
In the following sections we will present further extensive quantitative support for this scenario.

\subsection{Minimum excess energy density}

In Figs.~\ref{fig02}(a,b) we present the velocity scaling of the minimum energy, $\Delta E_{\rm min}$, defined for each disorder sample as 
the lowest energy reached at $T=0$ among any of the $32$ replicas run in parallel. We fix the exponent $z=13.6$ to be the same 
as that for the average energy shown in Fig.~\ref{fig01}. We see that the scaling also works very well here. If we instead treat the exponent as a variable and 
optimize its value for the best data collapse, we obtain $z=13.5(5)$ in excellent agreement (within the error bars) with the one previously obtained. Thus, as
expected, the two energies scale in the same way and the agreement also serves as a consistency check on the procedures. Note that, although the dynamic exponent 
is the same, the scaling functions are clearly different. In Fig.~\ref{fig02}(b) we plot out the two scaling functions in the same graph by only showing the data 
points that fall clearly on the collapsed curve. Given how the quantities are measured, at a given velocity, the minimum energy reached is always lower than 
(or in some cases equal to) the average energy after the final MC step. Based on a rough estimation from the two curves, $\langle \Delta E_{\rm min}\rangle$ 
relaxes about $10^4$ times faster to the asymptotic minimum value than $\langle \Delta E\rangle$. However, for larger values of the scaled velocity, and for 
sufficiently large system sizes, we expect the two energies to converge to the same power-law behavior with the exponent given by Eq.~(\ref{x_E}), and we 
see indications of this convergence as well in Fig.~\ref{fig02}(b). We can also see that our results for $\Delta E_{\rm min}$ are consistent with the prefactor 
$a=1$ in the equilibrium size dependence, Eq.~(\ref{de}), as the scaled quantity is close to $1$ in the low-velocity limit (though $a$ may also be marginally above $1$).
 
\subsection{Order parameter}

\begin{figure}[t]
\center{\includegraphics[width=7cm, clip]{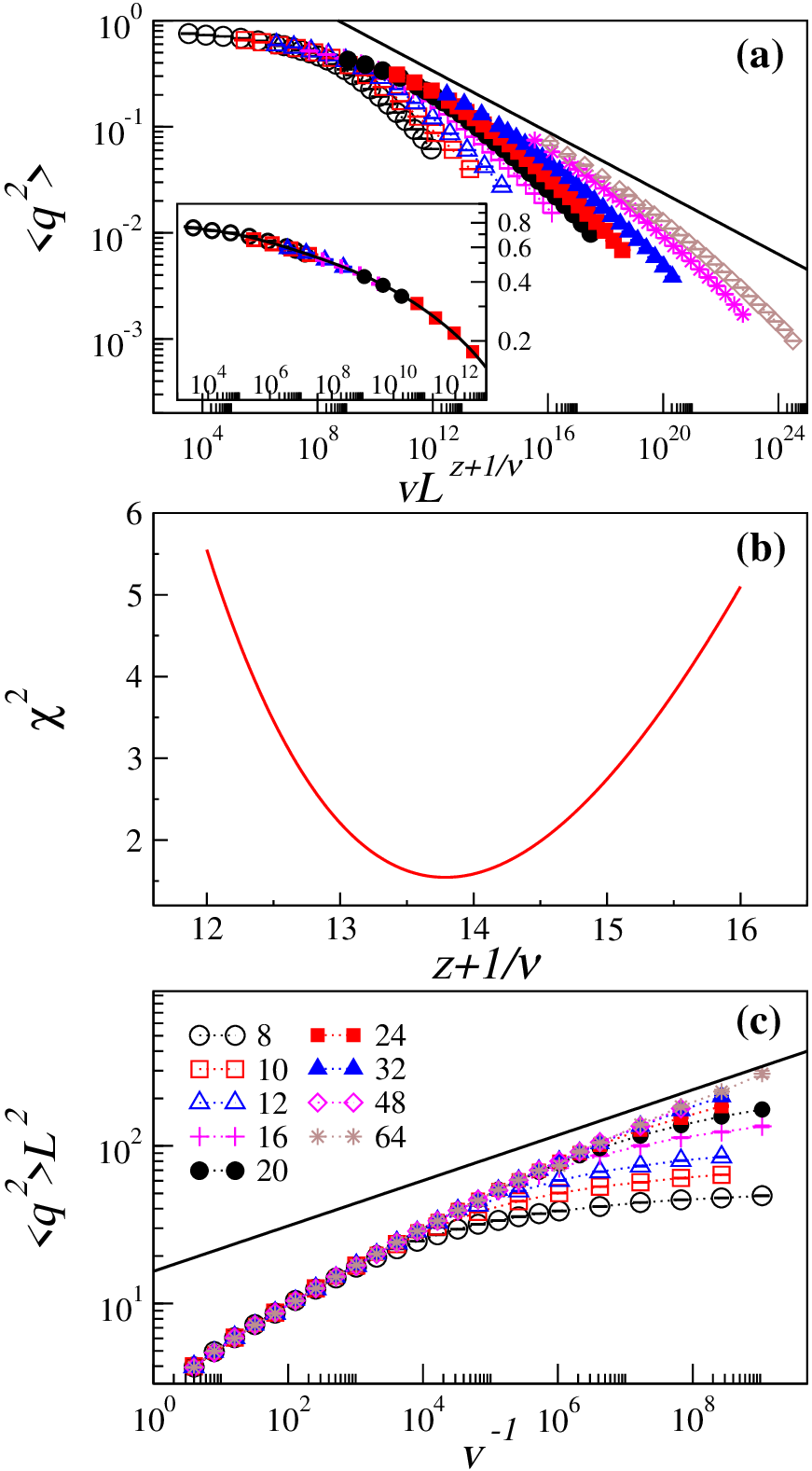}}
\vskip-1mm
\caption{Velocity scaling of the EA order parameter. In (a) the horizontal axis is rescaled according to the KZ ansatz with the dynamic 
exponent $z=13.6$ having the value extracted from $\Delta E$ in Fig.~\ref{fig01}. The straight line corresponds to the expected asymptotic
power-law behavior with the exponent $-x$ given in Eq.~(\ref{x_q}). To show more explicitly the quality of the collapse, the inset includes only the data
points used in the fitting procedure and the polynomial fitting function (black curve). Panel (b) shows the goodness of the fit, $\chi^2$ per degree of
freedom, versus the scaling exponent $z+1/\nu$. In (c) the data are graphed according to the third line of Eq.~(\ref{eq:q_scaling}), to show the non-universal
high-velocity behavior and its cross-over into the size-independent power-law behavior. The straight line has the same slope 
$x$ (up to the sign) as in (a).} 
\label{fig03}
\vskip-1mm
\end{figure}

\begin{figure}[t]
\center{\includegraphics[width=7cm, clip]{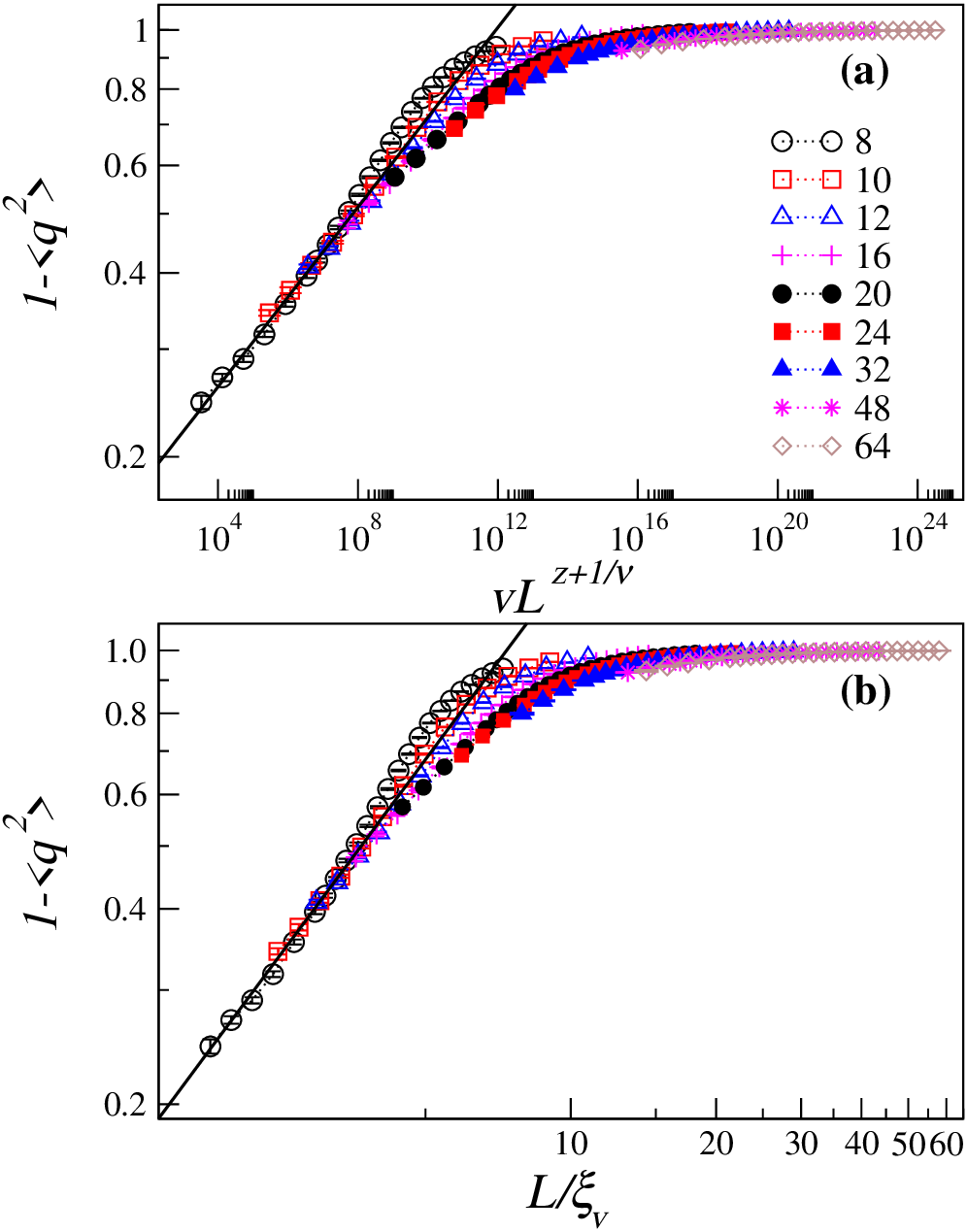}}
\vskip-1mm
\caption{
(a) The deviation $1-\langle q^2 \rangle$ from the asymptotic size-independent value $1$ graphed against the KZ-scaled velocity. The collapsed low-velocity
data are fitted to a power-law form (the line), $1-\langle q^2 \rangle \propto (vL^{z+1/\nu})^a$ with the exponent $a={0.073}$. (b) The same data as in (a) 
graphed against $L/\xi_v$, where the velocity-dependent correlation length is $\xi_v$ defined in Eq.~(\ref{velkz}) with the same exponent $z+1/\nu=13.9$
as in (a). The straight line here has slope exactly $1$.}
\vskip-1mm
\label{fig04}
\end{figure}

We next turn to the EA order parameter. Figures \ref{fig03}(a,b,c) show different aspects of the scaling of $\langle q^2\rangle$ with the velocity and the 
system size. In Fig.~\ref{fig03}(a), $\langle q^2 \rangle$ is graphed against the KZ-scaled velocity, using the same value of the dynamic exponent as was extracted
above using the excess energy. Here we cannot reach as close to the equilibrium behavior as for the energy (especially the minimum energy), but the approach 
of $\langle q^2\rangle$ to $1$ is still obvious and the data for the smaller system sizes collapse very well in this regime, as shown more clearly in the inset of 
Fig.~\ref{fig03}(a). The expected 
pure power-law behavior for large arguments $vL^{z+1/\nu}$ is not yet reached with the system sizes accessible here---the corrections to the power law as 
the equilibrium behavior is approached appear to be much larger than in the energy. The behavior is nevertheless consistent with an approach to the predicted 
asymptotic power-law scaling (indicated by the line in the figure). We also carried out the data collapse procedure with $z$ as a free parameter, using system 
sizes $L=8 - 24$ for which sufficient overlaps in the scaling variable exist so that the data-collapse procedure is well-defined. Figure \ref{fig03}(b) shows a
clear minimim in the $\chi^2$-value of the fit versus the scaling exponent, in very good agreement with the best exponent obtained for the energy scaling
in Fig.~\ref{fig01}. A full error analysis gives $z=13.6(2)$, which is consistent with but statistically better than $z=13.6(4)$ from the energy.
Thus, in contrast to the bimodal 2DISG, where a difference in dynamic exponents for the two quantities was found to be $z_E-z_q \approx 2$ ($z_E\approx 10.3$
and $z_q\approx 8.3$) \cite{rubin17}, in this case a single exponent governs the relaxation dynamics (as we had fully expected for this case where the
ground state is unique).

In Fig.~\ref{fig03}(c) we analyze the high-velocity limit of the order parameter, which eventually should cross over into the power-law regime. Recall that the 
collapse of data graphed versus the velocity (here the inverse velocity) at high velocities is trivial, merely reflecting the correlation length at the end of the 
SA process being much less than the system size (in the limit of $v \to \infty$ simply being the correlation length of the starting high-temperature equilibrium state), 
so that there is no size dependence. The initial state determines the details of the corresponding function $f_1(1/v)$ on the third line of Eq.~(\ref{eq:q_scaling})
at high velocities, before the cross-over into the universal form written explicitly on the second line. Here again, we see a very slow approach to the pure power 
law, similar to the cross-over from the low-velocity side, and we can only say that the behavior is consistent with the expected behavior with $z \approx 13.6$.

To investigate the approach to equilibrium in more detail, in Fig.~\ref{fig04}(a) we analyze the deviation $1-\langle q^2\rangle$ of the EA order parameter
from the asymptotic size-independent equilibrium value $1$. Here again we see good data collapse setting in from the left side of the graph and extending 
further to the right with increasing system size. In the region where $1-\langle q^2\rangle$ is small, the behavior follows a power law with a small, non-integer 
exponent. Here one would normally expect an integer exponent, corresponding to an analytic function $f_0(v/v_{\rm KZ})=f_0(vL^{z+1/\nu})$ on the first line of 
Eq.~(\ref{eq:q_scaling}). This has been observed in KZ scaling studies of non-random isolated quantum systems under Hamiltonian dynamics \cite{degrandi11},
for which the leading power laws for different quantities were also derived using adiabatic perturbation theory. Here the value of the exponent $a \approx 0.073$ 
in the power law $(L^{z+1/\nu})^a$ is very close to half of the value of the exponent $x$ in Eq.~(\ref{x_q}). Assuming
that $a=x/2=(z+1/\nu)^{-1}$, we see that the asymptotic form is
\begin{equation}
\langle q^2\rangle = 1 - bL/\xi_v,~~~~(L/\xi_v \to 0),
\label{q2new}
\end{equation}
where $\xi_v$ is the KZ correlation length corresponding to finite velocity in the thermodynamic limit \cite{polkovnikov05,zurek05,chandran12};
\begin{equation}
\xi_v \propto v^{-1/(z+1/\nu)},
\label{velkz}
\end{equation}
which can also be simply obtained from Eq.~(\ref{vkzdef}) by replacing $L$ by $\xi_v$.
Thus, we conclude that, unlike other cases studied so far \cite{degrandi11,liu14}, here $f_0(vL^{z+1/\nu})$ is not Taylor-expandable but a
corresponding function $\tilde f_0(L/\xi_v)$ is. We do not have an explanation for this apparently different analytic form of the scaling 
function in this case, but empirically the evidence is compelling, as seen more directly in Fig.~\ref{fig04}(b) where we plot the data against
$L/\xi_v$ and compare with a power-law with exponent exactly $1$, i.e., testing the asymptotic form Eq.~(\ref{q2new}).

One might perhaps question the claim that the observed power-law behavior in Fig.~\ref{fig04} should reflect the true asymptotic form, given that 
the scaling variable $vL^{z+1/\nu}$ is still very large in this region, roughly in the range $10^4-10^8$ in the power-law region.  However, the alternative
scaling variable $L/\xi_v$ is much smaller, of the order $1$. Since a scaling variable is always determined only up to some essentially arbitrary factor, a
more relevant measure of closeness to the asymptotic behavior should be the value of the quantity studied. Considering that $\langle q^2\rangle$ is as large 
as $0.8$, or, in other words, in two typical replicas $\approx 90\%$ of the spins are the same, and approximately the same fraction of the spins should 
then be in their ground-state configurations. We would then expect that the remaining relaxation of a dilute concentration of spins should already be 
governed by the asymptotic form, although we cannot completely exclude a cross-over into a different form still closer to equilibrium.
As we will see below, we can push a bit further into the low-velocity regime by considering smaller system sizes.

\begin{figure}[t]
\center{\includegraphics[width=7cm, clip]{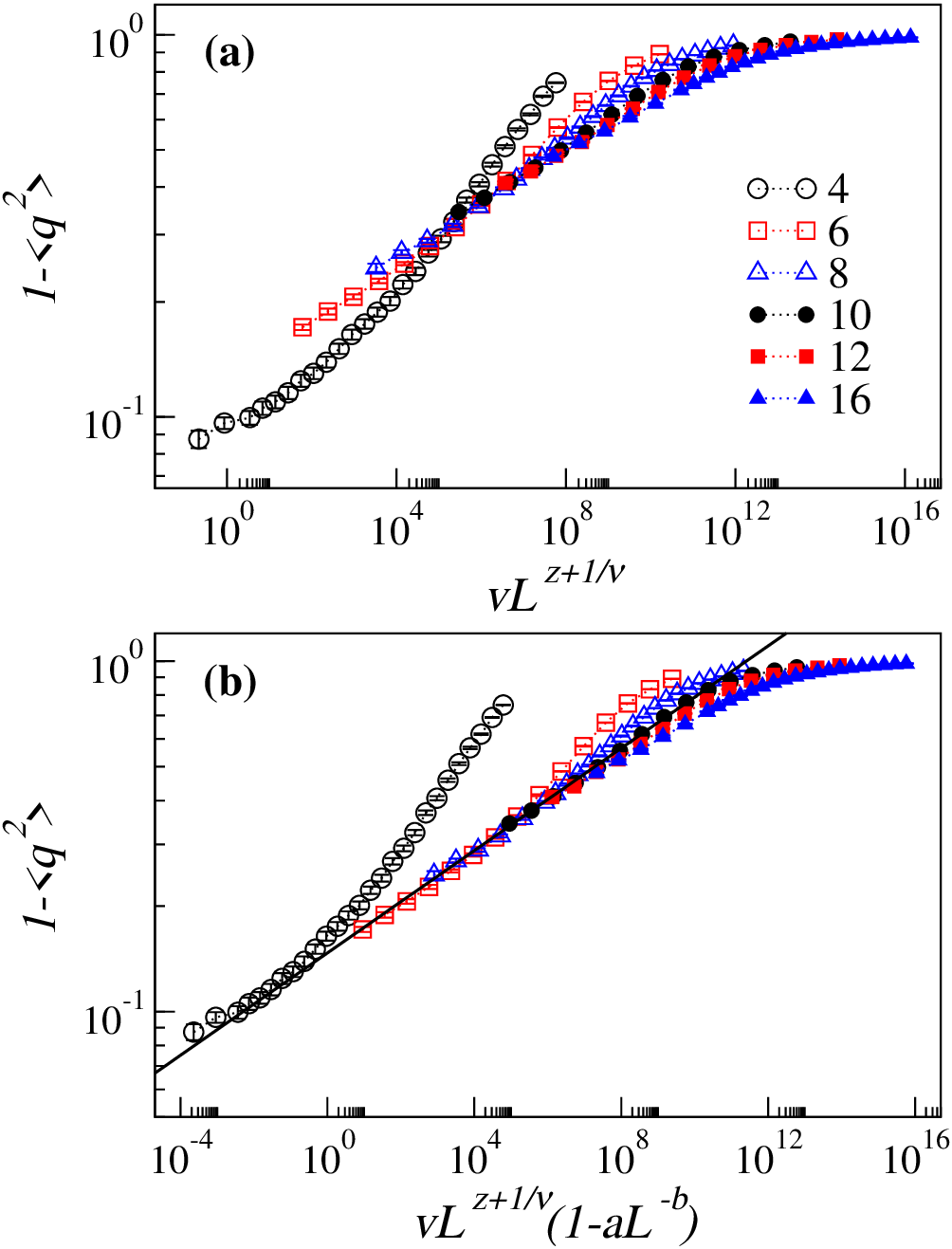}}
\vskip-1mm
\caption{Scaling of the deviation $1-\langle q^2 \rangle$ of the EA order parameter from its size-independent equilibrium value $1$, showing results only 
for small system sizes. In (a) the velocity is scaled according to the standard KZ form; the same as in  Fig.~\ref{fig04}(a). In (b) the scaling argument 
$vL^{z+1/\nu}(1-aL^{-b})$ contains a correction, with optimized parameter values $a = 1.7$ and $b= 0.39$. The line has the same slope as in Fig.~\ref{fig04}(a).}
\label{fig05}
\vskip-1mm
\end{figure}

In the above analysis of the EA order parameter, the smallest system size used in Figs.~\ref{fig03} and \ref{fig04} was $L=8$. For smaller sizes we see behaviors that
can be explained only with substantial scaling corrections included. Figure~\ref{fig05} focuses on the scaling of $1-\langle q^2 \rangle$ for small system 
sizes, from $L=4$ to $L=16$. In Fig.~\ref{fig05}(a), even though the $L\ge 8$ data collapse well in a region of slow velocities with standard KZ scaling and the
same value of $z$ used above, the data for $L=4$ and $L=6$ clearly deviate substantially from a common scaling function. Staying within the subset of
possible scaling corrections with no velocity dependence, we add a correction to the KZ argument $vL^{z+1/\nu}$ by multiplying it with $1-aL^{-b}$, 
with $a$ and $b$ optimized for the best data collapse (keeping $z$ at the previous value). With $a\approx 1.7$ and $b\approx 0.4$, the data collapse is very 
good on the left side, where also the power-law behavior found previously is substantially extended, with no detectable change in the exponent. This
gives added support to the power-law form corresponding to Eq.~(\ref{q2new}) indeed being the asymptotic behavior.

We have also tried to analyze the asymptotic approach of the energy density to its equilibrium value. Here we can in principle use the KZ ansatz following
from the known equilibrium finite-size scaling form Eq.~(\ref{de}) written in the following way:
\begin{eqnarray}
E(v,L) &=& E_\infty + aL^{-(2+1/\nu)}f(vL^{z+1/\nu}) \\
&=& E(0,L) + aL^{-(2+1/\nu)}g(vL^{z+1/\nu}),\nonumber
\end{eqnarray}
where $f(x) \to 1$ when $x=vL^{z+1/\nu} \to 0$ and $g(x) \to 0$ in this limit. Using the form of the equilibrium value, $E(0,L)=E_\infty + aL^{-(2+1/\nu)}$, 
with the parameters determined previously \cite{campbell04}, as mentioned below Eq.~(\ref{de}), we can analyze $(E(v,L)-E(0,L))L^{2+1/\nu}$. Within the standard 
scenario it should be a Taylor-expandable function $g(x)$ without constant term for small values of $x$. Unfortunately, here our results from Fig.~\ref{fig01} (from 
which we just need to subtract $1$ if the factor $a$ above really is exactly $1$, which is certainly consistent with our data in Fig.~\ref{fig02}) are not good enough 
(the statistical errors are too large) to extract any meaningful behavior in the low-velocity limit. We can therefore at present not determine whether an integer
power in $x$ obtains, or whether the leading behavior is instead an integer power of $L/\xi_v$ as in the case of $1-\langle q^2\rangle$.

\section{Discussion}
\label{sec:dis}

We have studied relaxation dynamics in the 2DISG model with Gaussian-distributed couplings by carrying out SA simulations in the $T\rightarrow 0$ limit, where
the system in equilibrium goes through a phase transition into the glass state. Through performing scaling analysis according to the KZ hypothesis, we were able 
to extract the dynamical exponents associated with the excess energy $\langle \Delta E\rangle$ and the EA order parameter $\langle q^2\rangle $. 

For the excess energy density, defined with respect to a previously determined value in the thermodynamic limit \cite{campbell04}, a data-collapse analysis yields 
$z=13.6(4)$, and the same kind of scaling procedure applied to the order parameter gives $z=13.6(2)$. Thus, there is a unique time scale governing the relaxation 
of both the order parameter and the excess energy. This in itself is not unexpected (as long as one accepts that the KZ mechanism applies),
but it is interesting in light of the recent discovery of two substantially different dynamic 
exponents in the 2DISG with bimodal couplings \cite{rubin17}. The heuristic explanation provided for that behavior relied on the massive degeneracy of the ground 
state, which is lacking in the case of couplings drawn from a continuous distribution. The ground state degeneracy has consequences for the relaxation of the mean 
order parameter as defined using replica overlaps. Considering that we here used the exact same kind of scaling procedures, our results for the Gaussian 
couplings also lend further support to the anomalous behavior in the bimodal case and its explanation in terms of ground-state degeneracy. 

The dynamic exponent we find here for the system with Gaussian couplings is significantly larger than the two different exponents for the bimodal couplings, 
where the larger of the two dynamic exponent, i.e., the one governing the energy relaxation, is $z'\approx 10.4$. While we do not have a rigorous explanation 
for this difference, it should be related to the fact that the ground state in the case of the bimodal couplings is degenerate, and, therefore, the process
does not have to find a specific unique spin configuration but is entropically attracted to a region with exponentially many ground states and is relaxed 
once any one out of these many configurations has been reached. 

Our results also further reinforce the notion that the relaxation dynamics of SA at these $T=0$ phase transitions is very different from the equilibrium 
dynamics, where it is known that, with local updates, the exponent governing the ergodic sampling process at fixed finite temperature diverges, 
$z_{\rm eq}(T)\rightarrow \infty$, when $T \to 0$  \cite{liang92,wang88,katzgraber05}. In contrast, at $T>0$ transitions, in both nonrandom and spin-glass 
models \cite{liu14,liu15a,fernandez16}, the dynamic exponent is finite and takes the same value at equilibrium and in SA analyzed within the KZ hypothesis. 
Clearly the source of this difference lays in the fact that the equilibrium dynamics is nonergodic in the limit $T \to 0$. 

Though the numerical evidence for KZ scaling of the SA dynamics is very strong, we do not have a rigorous theoretical explanation for why it applies,
instead of some exponentially slow relaxation dynamics related to naively expected activated scaling. The fact that power-law scaling does hold, in the
model studied here as well as in the previously studied case with bimodal couplings \cite{rubin17}, must reflect a
certain ``funnel'' structure of the energy landscape where the energy and entropy barriers along the walls down to the global minimum increase 
sufficiently slowly with the system size. This should be a consequence of the droplet picture in the model with bimodal couplings \cite{thomas11}, and also in the 
case of Gaussian couplings one can construct a similar approximate droplet structure \cite{roma10} that may explain the behavior found here. 

Given that KZ scaling in the form of data collapse onto a common scaling function is observed, a surprising behavior found here for the Gaussian couplings is 
that the scaling function for the EA order parameter does not appear to have a power-series expansion for small values of the standard KZ variable 
$vL^{z+1/\nu}$; instead the data show that the the scaling function has a Taylor expansion in the related variable $Lv^{1/(z+1/\nu)}=L/\xi_v$. This 
indicates a break-down of standard perturbative mechanisms behind KZ scaling in the low-velocity limit, which have been worked out for quantum many-body 
systems under Hamiltonian dynamics (quantum annealing) \cite{degrandi11} and have been shown to be applicable also for stochastic SA dynamics of classical 
systems \cite{liu14}. While the reasons for the non-perturbative behavior found here are not presently clear and deserve further study, one possibility is the 
proliferation of excited states nearly degenerate with the unique ground state, which may shrink the radius of convergence of the perturbation series to zero 
in the thermodynamic limit. How these non-perturbative effects lead to analytic behavior in the new scaling argument $L/\xi_v$ is not clear and is an important 
question for further study. Our result for the excess energy are not sufficiently precise to analyze the low-velocity corrections in that case. 

\begin{acknowledgments}
We thank David Huse and Anatoli Polkovnikov for helpful discussions. This work was supported by the NSF under Grant No.~DMR-1410126 (NX, SJR, and AWS), 
by MOST in Taiwan through Grants No.~104-2112-M-002-022-356-MY3 and 105-2112-M-002-023-MY3 (KHW and YJK), and by Boston University's Undergraduate 
Research Opportunities Program (SJR). Most of the computations were done on the National Center for High-performance Computing's Formosa 5 Cluster (Taiwan) 
and some of them were carried out on Boston University's Shared Computing Cluster.
\end{acknowledgments}


\begin{thebibliography}{99}

\bibitem{edwards75}
S. F. Edwards and P. W. Anderson, J. Phys. F: Met. Phys. {\bf 5}, 965 (1975).
 
\bibitem{binder86}
K. Binder and A. P. Young, Rev. Mod. Phys. {\bf 58}, 801 (1986).
 
\bibitem{fisher91}
K. Fisher and J. Hertz, Spin Glasses (Cambridge University Press, Cambridge, England, 1991).
 
\bibitem{parisi83}
G. Parisi, Phys. Rev. Lett. {\bf 50} 1946 (1983).
 
\bibitem{kawashima92}
N. Kawashima, N. Hatano, and M. Suzuki, J. Phys. A: Math. Gen. {\bf 25} 4985 (1992). 

\bibitem{campbell04}
I. A. Campbell, A. K. Hartmann, and H. G. Katzgraber, Phys. Rev. B {\bf 70}, 054429 (2004).

 \bibitem{houdayer04}
J. Houdayer and A. K. Hartmann, Phys. Rev. B {\bf 70}, 014418 (2004).

\bibitem{rieger96}
H. Rieger, L. Santen, U. Blasum, M. Diehl, M. J\"unger, and G. Rinaldi, J. Phys. A: Math. Gen. {\bf 29}, 3939 (1996). 

 \bibitem{jorg06}
T. J\"org, J. Lukic, E. Marinari, and O. C. Martin, Phys. Rev. Lett. {\bf 96}, 237205 (2006).

\bibitem{toldin11}
F. Parisen Toldin, A. Pelissetto, and E. Vicari, Phys.Rev. E {\bf 84}, 051116 (2011).

\bibitem{thomas11}
C. K. Thomas, D. A. Huse, and A. A. Middleton, Phys. Rev. Lett. {\bf 107}, 047203 (2011).
 
\bibitem{fernandez16}
L. A. Fernandez, E. Marinari, V. Martin-Mayor, G. Parisi, and J. J. Ruiz-Lorenzo, Phys. Rev. B {\bf 94}, 024402 (2016).

\bibitem{campbell16}
P. H. Lundow and I. A. Campbell, Phys. Rev. E {\bf 93}, 022119 (2016).

\bibitem{campbell17}
P. H. Lundow and I. A. Campbell, Phys. Rev. E {\bf 95},042107 (2017).

\bibitem{kibble76}
T. W. B. Kibble, J. Phys. A {\bf 9}, 1387 (1976).

\bibitem{zurek85}
W. H. Zurek, Nature (London) {\bf 317}, 505 (1985).

\bibitem{polkovnikov05}
A. Polkovnikov, Phys. Rev. B {\bf 72}, 161201(R) (2005).

\bibitem{zurek05}
W. H. Zurek, U. Dorner, and P. Zoller, Phys. Rev. Lett. {\bf 95}, 105701 (2005).

\bibitem{chandran12}
A. Chandran, A. Erez, S. S. Gubser, and S. L. Sondhi, Phys. Rev. B {\bf 86}, 064304 (2012).

\bibitem{liu14}
C.-W. Liu, A. Polkovnikov, and A. W. Sandvik, Phys. Rev. B {\bf 89}, 054307 (2014).

\bibitem{liu15a}
C.-W. Liu, A. Polkovnikov, and A. W. Sandvik, Phys. Rev. Lett. {\bf 114}, 147203 (2015).

\bibitem{zhong05}
F. Zhong and Z. Xu, Phys. Rev. B {\bf 71}, 132402 (2005).

\bibitem{liu15}
C.-W. Liu, A. Polkovnikov, A. W. Sandvik, and A. P. Young,  Phys. Rev. E {\bf 92}, 022128 (2015).

\bibitem{rubin17}
S. J. Rubin, N. Xu, and A. W. Sandvik, Phys. Rev. E {\bf 95}, 052133 (2017).

\bibitem{roma10}
F. Rom\'a, S. Bustingorry, and P. M. Gleiser, Phys. Rev. B {\bf 81}, 104412 (2010).

\bibitem{katzgraber15}
H. G. Katzgraber, F. Hamze, Z. Zhu, A. J. Ochoa, and H. Munoz-Bauza, Phys. Rev. X {\bf 5}, 031026 (2015).

\bibitem{katzgraber05}
H. G. Katzgraber and I. A. Campbell, Phys. Rev. B {\bf 72}, 014462 (2005).

\bibitem{roma13}
F. Rom\'a and S. Risau-Gusman, Phys. Rev. E {\bf 88}, 042105 (2013).

\bibitem{roma16}
F. Rom\'a, S. Bustingorry, and P. M. Gleiser, Eur. Phys. J. B  {\bf 89}， 259 (2016).

\bibitem{thomas07}
C. K. Thomas and A. A. Middleton, Phys. Rev. B {\bf 76}, 220406(R) (2007). 

\bibitem{preis09}
T. Preis, P. Virnau, W. Paul, and J. J. Schneider, J. Comp. Phys. {\bf 228}, 4468 (2009).

\bibitem{block10}
B. Block, P. Virnau, and T. Preis, Comp. Phys. Comm. {\bf 181}, 1549 (2010).

\bibitem{lulli15}
M. Lulli, M. Bernaschi, and G. Parisi, Comp. Phys. Comm. {\bf 196}, 290 (2015).

\bibitem{hsieh13}
Y.-D. Hsieh, Y.-J. Kao, and A. W. Sandvik, J. Stat. Mech. {\bf 2013}, P09001 (2013).

\bibitem{hartmann15}
M. Manssen and A. K. Hartmann, Phys. Rev. B {\bf 91}, 174433 (2015).

\bibitem{fang14}
Y. Fang, S. Feng, K.-M. Tam, Z. Yun, J. Moreno, J. Ramanujam, and M. Jarrell, Computer Physics Communications {\bf 185}, 2467 (2014).

\bibitem{jesi14}
M. Baity-Jesi, L. A. Fern\'andez, V. Mart\'in-Mayor, and J. M. Sanz, Phys. Rev. B {\bf 89}, 014202 (2014).

\bibitem{parisi10}
L. P. M. Bernaschi, G. Parisi, arxiv: {\bf 1006.2566} (2010).

\bibitem{weigel12a}
T. Yavors'kii and M. Weigel, The European Physical Journal Special Topics {\bf 210}, 159 (2012).

\bibitem{weigel12b}
M. Weigel, Journal of Computational Physics {\bf 231}, 3064 (2012).

\bibitem{hartmann11}
A. K. Hartmann, J. Stat. Phys. {\bf 144}, 519 (2011).

\bibitem{liang92}
S. Liang, Phys. Rev. Lett. {\bf 69}, 2145 (1992).

\bibitem{wang88}
J.-S. Wang and R. H. Swendsen, Phys. Rev. B {\bf 38}, 4840 (1988).

\bibitem{degrandi11}
C. De Grandi, A. Polkovnikov, and A. W. Sandvik, Phys. Rev.  B {\bf 84}, 224303 (2011).



\end{thebibliography}
\end{document}